\documentclass[pre,preprint,aps,]{revtex4}
\usepackage{subcaption}
\usepackage{graphicx}
\usepackage{comment}
\usepackage{amsmath,amssymb,amsthm} 
\usepackage{bm}
\usepackage{verbatim}
\usepackage{float}

\begin{document}
\title[Planetary systems with forces other than gravitational forces ]{Planetary systems with forces other than gravitational forces }
\author{ S\o ren  Toxvaerd }\email{st@ruc.dk}
\affiliation{ Department
 of Science and Environment, Roskilde University, Postbox 260, DK-4000 Roskilde, Denmark}
\date{\today}

\vspace*{0.7cm}

\begin{abstract} A discrete and exact algorithm for obtaining planetary systems is derived in a recent
		article (Eur. Phys. J. Plus 2022, 137:99). Here the algorithm
			is used  to obtain planetary systems with forces different from the Newtonian inverse square gravitational forces.
			A Newtonian planetary system exhibits regular elliptical orbits, and here it is demonstrated that a planetary system with pure inverse forces
		also	is stable and with
			 regular orbits, whereas a planetary system with inverse cubic forces is unstable and without regular orbits. The regular orbits in
			a planetary system with inverse forces deviate, however, from the usual elliptical orbits by having revolving orbits with tendency to
			orbits with three  or eight loops. Newton's Proposition 45 in $\textit{Principia}$ for the Moon's revolving orbits caused by an
				additional attraction to the gravitational attraction  is confirmed, but whereas the additional  inverse forces stabilize the
				planetary system, the additional  inverse cubic forces can destabilize the planetary system at a sufficient strength.
\end{abstract}				


\maketitle

\section{Introduction}

Our world consists of objects with collections of   atoms and molecules which are bound together by ionic or covalent bonds.
On a larger length scale
these objects are collected in planetary systems in galaxies, which are bound together by gravitational forces. The ionic and covalent bonds are
established by electromagnetic forces whereas the planetary systems and the galaxies are hold together by
gravitational forces. Although the two forces differ enormously in strength
by a factor of $\approx 10^{36}$ they have, however, some  common features. The
radial strengths of both forces are proportional to the inverse square (ISF),  $r^{-2}$,  of the  distances between mass centers , and
both forces are believed to extend to infinity.
 The two forces can also result in regular closed orbits for  the dynamics of a collection of force centres, as is
demonstrated by our solar system and the orbitals of the bounded electrons at a
atomic nucleus. The two other fundamental forces are the strong and weak nuclear forces and
they are both short ranged.  All other forces are ''derived forces" such as the 
harmonic forces or the attractive induced dipole-dipole forces.

 Isaac Newton formulated the classical mechanics in his book  PHILOSOPHI\AE \ NATURALIS PRINCIPIA MATHEMATICA ($Principia$)  \cite{Newton1687},
  where he also proposed the law of gravity and solved  Kepler's equation   for a planets motion.  According to Newton,
   gravity varies with  the inverse square  of the distance $r$ between two celestial objects, and a planet exposed to the gravitational
    force from the Sun moves  in  an elliptical
     orbit. The Moon exhibits, however,  periodic
      ''revolving orbits" and Newton shows  in $Principia$ that this behaviour, which is caused by the daily rotation of the Earth, could
       be taken into account by and additional inverse cubic force proportional to $r^{-3}$ (ICF). But it raises the question: for  which
        forces can a system of objects   have regular orbits?

It is only possible to solve the  classical mechanics differential equations for two objects.
The  classical second-order
differential equation for the dynamics of  two objects with a central force proportional
to $r^n$ can be solved for a series of values of the power $n$ .
An important result was obtained by Bertrand \cite{Bertrand}, who  proved that all 
bound orbits  are closed orbits. Later investigations have proved the existence of regular orbits 
 for a series of values of the power $n$ of the central force, including the ICF \cite{Whittaker,Broucke1980,Mahomed2000}. 
 
For a system  consisting of many objects, the dynamics of   
 coupled  harmonic oscillators   demonstrates that it indeed is
 possible to have stable regular dynamics for systems with other forces than the gravitational forces,
but else there is no theoretical proofs, nor  any other  examples of that it
is possible.
Here it is, however, demonstrated by Molecular Dynamics simulations (MD) of planetary systems \cite{Toxvaerd2022}, that a planetary system also
can have planets with stable regular  orbits for attractive forces which
varies as\\
$-r^{-1}$\\
$-r^{-2} \pm \alpha \times r^{-1}$\\ 
$-r^{-2} \pm \alpha \times r^{-3}$ for $\alpha \in [-100,10]$.\\
But  it  has not been possible to obtain stable regular orbits  for  $r^{-3}$.

\section{ The force between two spherically symmetrical objects}\label{sec 2}

   Newton was aware of that the extension of an object 
  can affect the gravitational force between two objects, and in $\textit{Theorem XXXI} $ in $Principia$ \cite{Newtonshell}  
	  he also solved  this problem for  ISF between spherically symmetrical objects.
	  
Newton's  $\textit{Theorem XXXI} $  states that:

  1.  A spherically symmetrical body affects external objects gravitational as though all of its mass were concentrated at a point at its center.
  
  2.      If the body is a spherically symmetric shell no net gravitational force is exerted by the shell on
  any object inside, regardless of the object's location within the shell.

 Newtons theorem is, however, only valid for ISF. The forces between spherically symmetrical objects
with forces proportional to  $r^{-1}$, inverse forces (IF), or ICF depends on the objects extension.
Newton's derivation of the theorem is by the use of Euclidean geometry, but the forces between two spherically symmetrical objects
can also be derived by the use of  algebra.

Let the objects No. $i$ and  $j$ be spherically symmetrical with masses $m_i$ and $m_j$ and 
with a uniform  density within the balls with the radii $\sigma_i$ and $\sigma_j$.
The attraction, IF, ISF or ICF,  on a mass $\delta m_i$ at $\textbf{s}_i$  in object $i$ 
and at the distance $s_{ij}$ from a mass  $\delta m_j$  at $\textbf{s}_j$ in $j$ 
 is 
\begin{equation}
	\delta \mathbf{F}_{ij}=-\beta \delta m_i \delta m_j s_{ij}^n \hat{\mathbf{s}}_{ij},	
\end{equation}
with $n$=-1, -2 and -3, respectively,
and
the  total force $\textbf{F}_{ij}$ is obtained by a quadruple  integration, first between $\delta m_i$  at $\textbf{s}_i$ and
 mass elements $\delta m_j$   in a sphere in $j$ with radius $\sigma_j^{'} \le\sigma_j$,
then over spheres centred  at $\textbf{r}_j$ with radius  $\sigma_j^{'}$,
 and then correspondingly between mass $m_j$ located in the center, $\textbf{r}_j$ and mass elements $\delta m_i$.

Consider mass elements
$\delta m_j(\textbf{s}_j)=4 \pi \sigma_j^{'2} m_j d\sigma'_j/(4 \pi/3  \sigma_j^3)$ at $\textbf{s}_{j}$  in a
thin shell $[\sigma'_j,\sigma'_j+d \sigma'_j]$ with center at $\textbf{r}_j$ and a distance
	$r'_{ij}=\mid \textbf{s}_i- \textbf{r}_j \mid > \sigma_i+\sigma_j \ge \sigma_i+\sigma'_j$  to   $\textbf{s}_i$.
	The force $\delta \textbf{F}_{ij}=-\beta \delta m_i m_j s_{ij}^n\hat{\textbf{r}}'_{ij}$ on $\delta m_i$ from object $j$ is
  \cite{WikipediaNewtonshell} 
\begin{equation}
	\delta \mathbf{F}_{ij} =-\beta  \frac{\delta m_i}{4r_{ij}^{'2}} \int_0^{\sigma_j} \frac{\delta m_j}{\sigma'_j}\int^{r'_{ij}+
	\sigma'_j}_{r'_{ij}-\sigma'_j}s_{ij}^n[s_{ij}^2+r_{ij}^{'2}-\sigma_j^{'2}] ds_{ij}\hat{\mathbf{r}}'_{ij}.	
\end{equation}

 The integrals
are very simple for ISF since
\begin{eqnarray}
	-\beta  \frac{\delta m_i}{4r_{ij}^{'2}} \int_0^{\sigma_j} \frac{\delta m_j}{\sigma'_j}\int^{r'_{ij}+
		\sigma'_j}_{r'_{ij}-\sigma'_j}s_{ij}^{-2}[s_{ij}^2+r_{ij}^{'2}-\sigma_j^{'2}] ds_{ij}\\ \nonumber 
=	-\beta  \frac{\delta m_i}{4r_{ij}^{'2}} \int_0^{\sigma_j} \frac{\delta m_j}{\sigma'_j} 4\sigma'_j=-\beta \frac{\delta m_i m_j}{r_{ij}^{'2}},
\end{eqnarray}			
and the integration over shells centred at $\textbf{r}_i$ with mass elements $\delta m_i$ leads to
$Theorem$ $XXXI$.
		
The integrations are more complex for $n \ne -2$. The simplest way to proceed is to expand the first integral in powers of
$\sigma'_j/r'_{ij}$. The first terms in the final expressions for the  force between $i$ and $j$ are given below. 

For the IF function $s^{-1}$:
\begin{equation}
	\mathbf{F}_{ij}(r_{ij}) \simeq -\frac{\beta_1 m_im_j}{r_{ij}}(1 -\frac{\sigma_i^2+\sigma_j^2}{5r_{ij}^2})\hat{\mathbf{r}}_{ij}+ \mathcal{O}(r_{ij}^{-4}) 
\end{equation}
For  $s^{-2}$ one obtains the usual expression for the gravitational ISF force ($\beta_2 = G$) which does not depend on the extensions of the two spherically symmetrical objects
\begin{equation}
	\mathbf{F}_{ij}(r_{ij})  = -\frac{G m_im_j}{r_{ij}^2}\hat{\mathbf{r}}_{ij} .	
\end{equation}
For  $s^{-3}$   the ICF radial force is
\begin{equation}
	\mathbf{F}_{ij}(r_{ij}) = -\frac{\beta_3 m_im_j}{r_{ij}^3}(1 +\frac{2\sigma_i^2+2\sigma_j^2}{5r_{ij}^2})\hat{\mathbf{r}}_{ij}+ \mathcal{O}(r_{ij}^{-6}). 
\end{equation}

The  dynamics of planetary systems with the different kind of
gravitational attractions is given in the next Section.

\begin{figure}
\begin{center}	
\includegraphics[width=7cm,angle=-90]{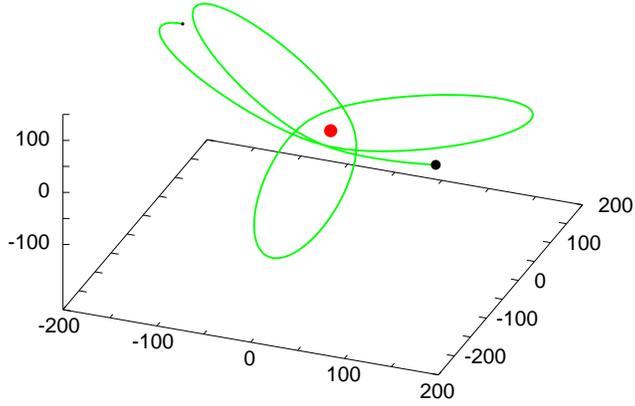}
\caption{A  loop of the innermost planet  from a position at time $t=2.5\times 10^6$, marked 
	by a big black sphere to a position at $t=2.5007325\times 10^6$ (293000 discrete time steps),  marked by a small black sphere. 
	The position of the ''Sun" is with an enlarged red sphere. Some simultaneous loops of two other
	planets in the planetary system are shown in the next figure.}
\end{center}
\end{figure}	
\begin{figure}
\begin{center}	
\includegraphics[width=7cm,angle=-90]{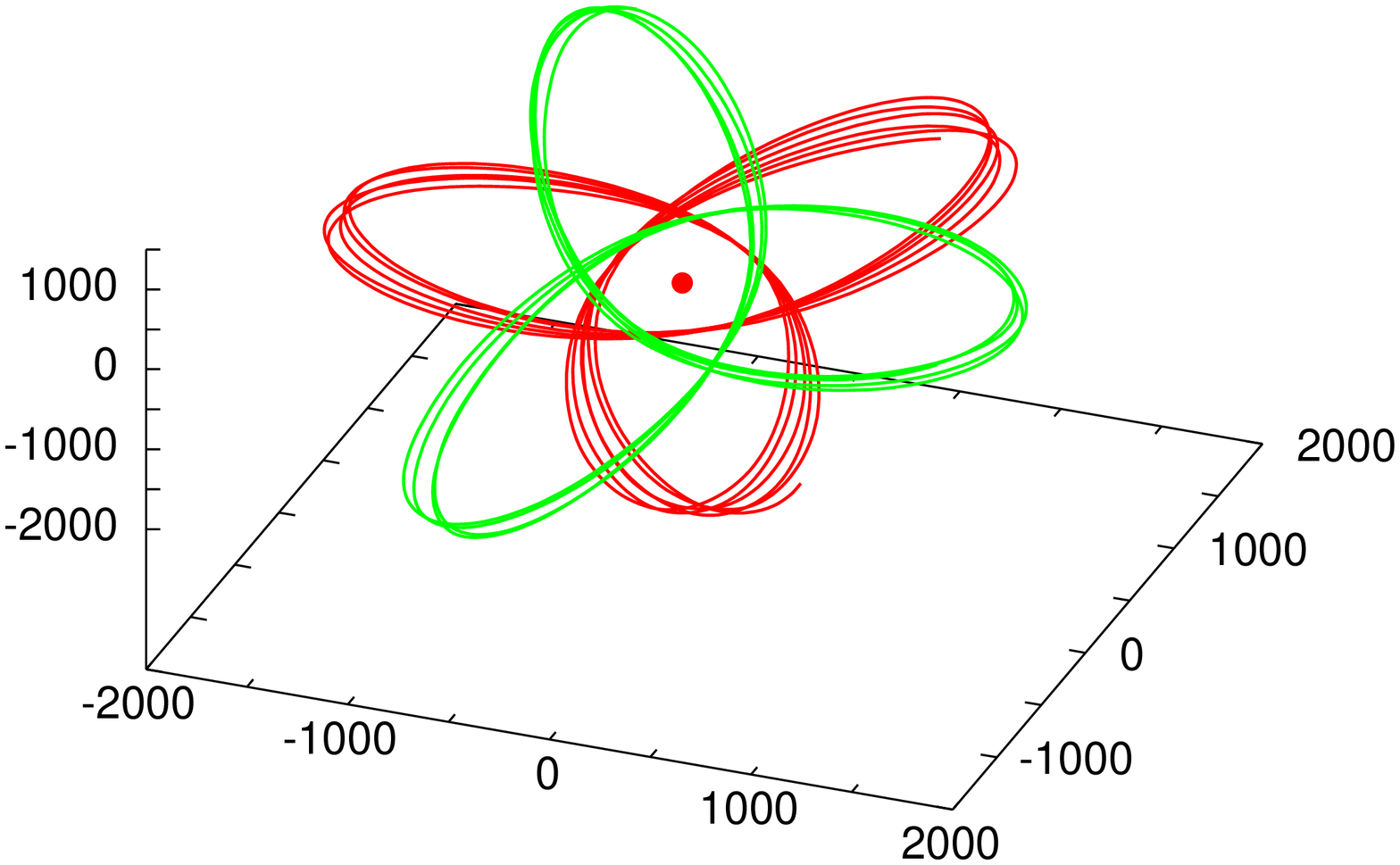}			
\caption{ The simultaneous orbits with bows for two other planets in the planetary system. The orbits are obtained for one million discrete time steps in  the time interval
$t \in [2.5\times 10^6, 2.5025\times 10^6]$.}
\end{center}

\end{figure}

\section{Dynamics of planetary systems  with different gravitational forces}
A discrete and exact algorithm for obtaining planetary systems is derived in a recent
		article \cite{Toxvaerd2022}. The algorithm  is symplectic and time reversible and has the
		same invariances as Newton's analytic dynamics. For Kepler's solution of the two body system of a Sun and a planet one
		can compare the two dynamics \cite{Toxvaerd2020}, which leads to the same orbits. The discrete dynamics
		is absolute stable and without any adjustments for
		conservation of energy, momentum and angular momentum for billion of  time steps. The
		algorithm and how to obtain planetary systems is given in the Appendix. Here the algorithm
					is used to obtain planetary systems with forces different from the Newtonian inverse square gravitational forces.

\begin{figure}
	\begin{center}	
\includegraphics[width=7cm,angle=-90]{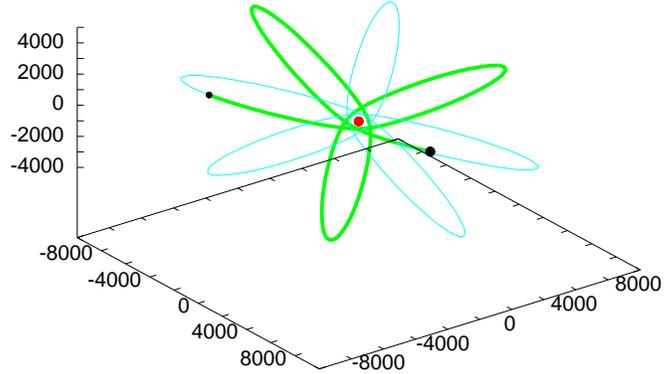}
\caption{ Bows in a loop with green and light blue
for the outermost planet. The
start position at $t=2.5\times 10^6$ is marked with a big black sphere and the first three
bows is with green color. The position at $t=2.5025\times 10^6$ after the first three bows is shown by a smaller black sphere,
and the succeeding five bows is with light blue.															
Several consecutive loops of the planet are shown in Figure 5.}
	\end{center}

\end{figure}

\begin{figure}
	\begin{center}	
	\includegraphics[width=7cm,angle=-90]{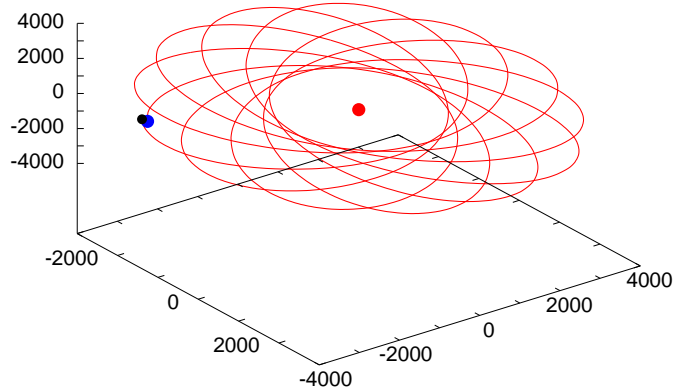}
	\caption{ A planet which after 13 orbits almost returns to its start position. The start position at $t=2.5\times 10^6$
		is shown with a big blue sphere, and the  position at  $t=2.5051975\times 10^6$		
	after thirteen orbits by a smaller black sphere.
	The next figure shows several orbits of the planet together with the orbits of the outermost planet.}	
	\end{center}

\end{figure}

\subsection{ Planetary systems  for objects with  inverse forces}

The IF between two objects $i$ and $j$ is given by the Eq. (4). The Eq. (4) gives the first-order size-correction     for the  forces
between spherically symmetrical uniform mass objects. The  investigation is conducted in two ways by MD simulations. $\textbf{A}$: One can simply create planetary systems
in the same way as described in \cite{Toxvaerd2022} and in the Appendix, or   alternatively  $\textbf{B}$: one can replace the Newtonian ISF  forces between objects in an ordinary planetary system
by the corresponding inverse IF forces.

 $\textbf{A}$: The results of obtaining planetary systems spontaneously by merging of objects as in \cite{Toxvaerd2022}
are shown in the next Figures. Planetary systems  with strength $\beta_1=1$ 
were created spontaneously   at time $t=0$   from  different  configurations, distributions of
velocities and     masses $m_i(0)=1$ 
of objects. (For units of length, time and  strength of the attractions in the MD systems  see the Appendix.) 
Ten different planetary systems were formed and
the overall result and conclusion from the ten systems is, that it is easily to obtain planetary systems with IF forces.
But the regular orbits
deviate, however, qualitatively from the elliptical orbits in an ordinary planetary system. A typical regular orbit is shown in Figure 1.

  Figure 1 shows a loop  of the innermost planet in one of the ten  planetary systems, which was simulated with IF. The planetary
  system was started  with thousand objects and the planetary  system with IF contained 38 planets after $10^9$ MD time steps corresponding
  to a MD time $t=2.5\times 10^6$, and where the inner planets have performed several thousand bound rotations. 
  The planet in Figure 1 performs
  a loop, but with a change of its elliptical major axis  by $\approx \pi/3$ at the passage of the ''Sun", by which the total regular orbit appears with three consecutive bows with 
   an angle of $\approx2\pi/3$.
  The total angular momentum for the system is conserved by Newton's exact discrete algorithm \cite{Toxvaerd2022}, but also the angular momentum of the individual planets in the system
  are conserved to a high degree so the three bows are in the same plane. Most of the planets exhibit this regular dynamics. Figure 2 shows
  the simultaneous orbits for two other planets in the same planetary system. The planetary system with the object shown in Figure 1 and Figure 2 consists of 38 objects
  in bound orbits around a central heavy object (the ''Sun" with $m_{\textrm{Sun}}$=867). In \cite{Broucke1980} Broucke has obtained the orbit for one planet (Figure 2 in  \cite{Broucke1980}).
  There is, however,  only some similarities between the present orbits for a many-body  three dimensional planetary system  and the 2D orbit of a single planet.

  \begin{figure}
\begin{center}	
 \includegraphics[width=7cm,angle=-90]{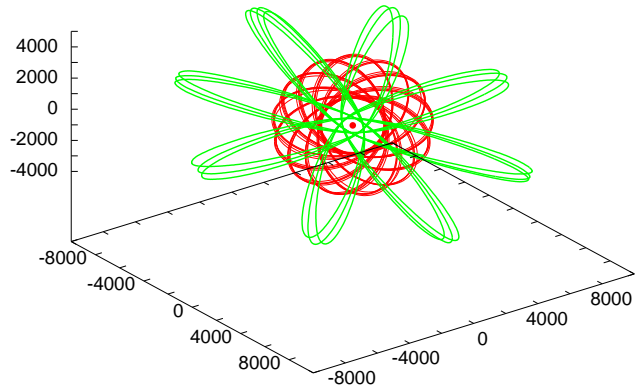}
\caption{ The outermost planet (green) together with the planet from Figure 4 (red) with its orbits in bands.  The outermost planet
 changes its major orbit  axis  with $\approx \pi/4$ at every past of the Sun.
 The central Sun is with red.}
  \end{center}

  \end{figure}

 All the planets in the  planetary systems with inverse forces show, what Newton probably would have called revolving orbits, but not all
 of the planets have  orbits of the form shown in Figure 1 and Figure 2. The next figure, Figure 3, show the consecutive bows of the outermost
 planet in the same planetary system. The planet changes its principal axis by $\approx \pi/4$ by which it performs eight bows in its regular orbit. Figure 5 shows with green 3-4 loops
 of this planet together with another planet in the same planetary system.

  It has not been possible to obtain simple elliptical regular orbits, but there are examples of planets with a smaller change of their principal axis
  at the passage of the Sun. Figure 4 gives such an example of a planet, which after thirteen loops return to its start position, and a collection of consecutive
  loops for this planet, shown by red in Figure 5, demonstrates that this regular pattern is maintained over many consecutive loops.

 $\textbf{B}$: Planetary systems with IF forces were obtained in another way by replacing the  Newtonian ISF forces
in an ordinary planetary system 
with the IF forces. The discrete dynamics with IF was started with the end-positions of the planet in the  planetary system \cite{Toxvaerd2022}.
A replacement with  $\beta_1=\beta_2=G$ 
results in a collapse of the planetary systems and with only two planets 
  in revolving orbits similar to the orbits shown in Figure 2. The other planets were engulfed by the Sun.
This is due to that the inverse forces  with  $\beta_1=\beta_2=G$ and acting on a planet are about a thousand time stronger than the Newtonian gravitational forces.
Planets in the Newtonian planetary systems in \cite{Toxvaerd2022} are  located at mean distances to their Suns at
$<r_{i,Sun}> \approx [100,30000]$. For an ordinary planet with a Newtonian ISF force field and  at a position  $r_{i,Sun}= 1000$
the corresponding IF force  is of the order  thousand time
stronger than the Newtonian ISF force. So in order to establish whether it is possible to obtain simple elliptical orbits without
revolving orbits, the forces in   the Newtonian  planetary systems  in \cite{Toxvaerd2022} were replaced with IF forces and with $\beta_1 \approx G/1000$.
The replacement was performed in the following way:

A planet $i$ with a rather circular orbit and at a mean distance  $<r_{i,Sun}> \approx 1000$ was selected and the strength $\beta_1=0.00105$ was
determined so the planet follow the same elliptical orbit shortly after the replacement. 
The result of this replacement of the forces in the planetary  system  on the orbit of this planet is  shown in Figure  6,
which show the orbit of an ordinary planet before( red) and after (green)
the replacement. The replacement is for $\beta=0.00105$ for which the planet followed the gravitational orbit (red) over a long period of time before
it deviated and exhibited the revolving orbits shown in the figure, but with a small change of its principal axis by passage at the Sun and with elliptical-like
orbits. The other planet in the Newtonian planetary system changed their orbits to the
  revolving orbits (blue orbit Figure 6), also shown in the previous figures.

 It has not been possible to obtain simple elliptical orbits, which spontaneously appears in an ordinary Newtonian planetary system. The simulations were performed by the
 first order IF expression, Eq. 4, but  simulations with- and without the first order correction
 $\mid\delta\textbf{F}_{\textrm{IF}}\mid= \beta_1  m_i m_j(\sigma_i^2+\sigma_j^2)/5r_{ij}^3$ showed, that the first-order correction  only has a minor 
 quantitative effect, and that the exclusion of this term do not change the overall qualitative result.

\subsection{Simulation of systems with inverse cubic forces}

\begin{figure}
	\begin{center}	
\includegraphics[width=7cm,angle=-90]{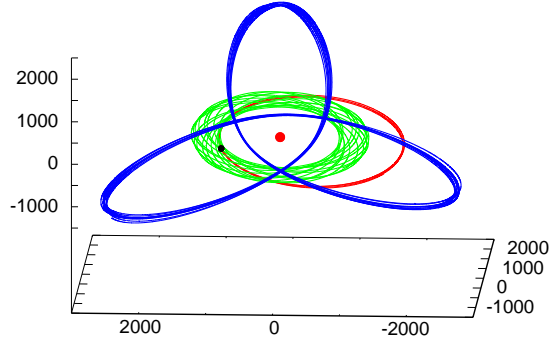}
\caption{ The red elliptical orbits is for a planet with Newtonian ISF forces and the green orbits are after the forces
at the position marked with a black sphere is replaced with IF forces and with $\beta_1$=0.00105 by which the planet
		in a short time follow the elliptical path  before the revolving behaviour.
		The orbit (blue) of a planet at a mean distance slightly bigger than the planet shown by red
changed spontaneously its elliptical orbit to the bows also shown in the previous figures.}
	\end{center}
	
\end{figure}

\begin{figure}
	\begin{center}	
\includegraphics[width=7cm,angle=-90]{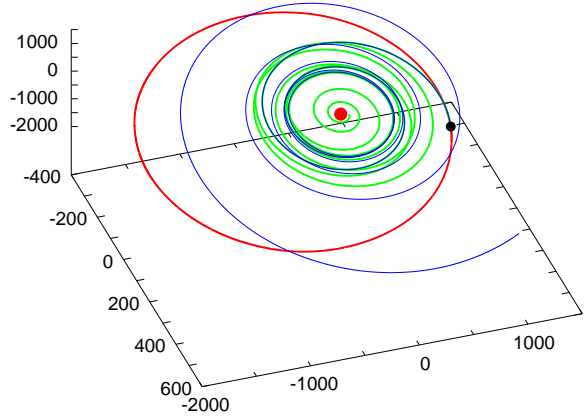}
\caption{ The orbits for a planet in a planetary system  with ICF.
The planetary system is obtained from an ordinary planetary system and with elliptical
orbits (red) by replacing the ISF forces by ICF and
with a strength $\beta_3 \approx 1230\times G$. The black circles is the position of the planet
at the time where the replacement took place and the green curve is for $\beta_3=1228.75$ and
the blue curve is for $\beta_3=1228.5$.}
	\end{center}

\end{figure}

\begin{figure}
	\begin{center}	
\includegraphics[width=7cm,angle=-90]{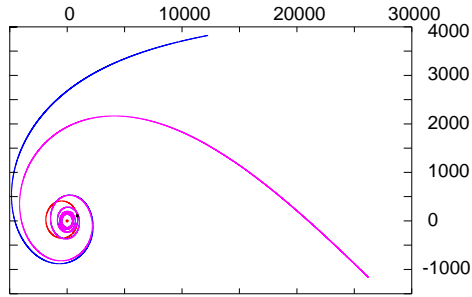}
\caption{Top view of the orbits of the planet also shown in details in the previous figure. The blue curve is for
	ICF with Eq. (6) and the magenta curve is with the zero-order ICF $-1228.5m_im_j/r_{ij}^3$.}
	\end{center}

\end{figure}

A system of objects with masses $m_i(0)=1$ and pure  ICF  given by Eq. (6) does not self-assemble to a planetary system. The objects either
fuse together or expand as free objects. This observation is valid for different values of the gravitational constant $\beta_3$ in Eq. (6) and it was
not possible to create a planetary system with ICF.

Another way to demonstrate the instability of planetary systems with pure ICF attractions is to replace the
Newtonian gravitational ISF in a planetary system by ICF as described in the previous subsection. Thus it is possible to determine a
values of $\beta_3 \pm \delta$, by which a given planet in an Newtonian planetary system either engulfs by the Sun by changing the forces from
$-G/r^2$ to $-(\beta_3+\delta)/r^3(1+(2\sigma_i^2+2\sigma_j^2)/5r_{ij}^2)$ or leaves the Sun
as a free object for $\beta_3-\delta$. Figure 7 shows this ''tipping point" for the same planetary system and planet as is shown in
Figure 6 with red for ISF and green for IF. The planet with pure ICF is engulfed by the Sun for $\beta_3+\delta=1228.75$ (green curve),
but escapes the Sun for $\beta_3-\delta=1228.5$ (blue curve). The Eq. (6) is with
the first asymptotic correction in a rapid converging expansion for the extension of the spherically symmetrical objects with
an uniform density. The zero order expression for the ICF system: $-\beta_3 m_im_j/r_{ij}^3$ gives the same
qualitatively result, as shown in Figure 8.  The tipping point is the same either one includes the first order correction or not.

  \section{ Newton's proportions for the Moon's revolving orbits}

\begin{figure}
	\begin{center}	
	\includegraphics[width=7cm,angle=-90]{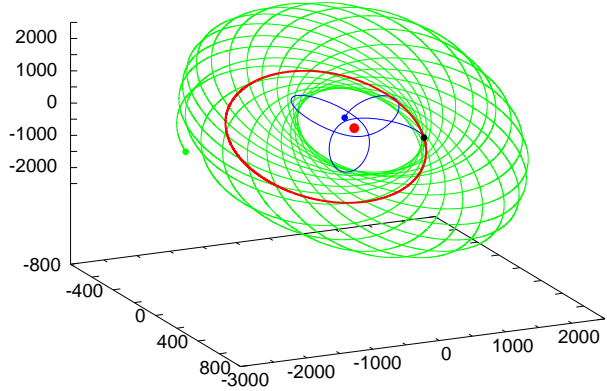}
	\caption{The planet also shown in Figure 7  with red, but now with IF or ICF included in the ISF attraction.
	 The orbit
	with red is with pure ISF; the orbit with green is with  ICF: $\alpha_3/r^3=-100/r^3$ included from the  position marked by a black sphere, and
 the orbit with blue is with the IF: $\alpha_1/r=0.01/r$ included.
	The  planet   with ISF + ICF (green) has revolved $\approx$ 23-24 times before
	the principal axis in the elliptical orbit has changed $2\pi$.}
	\end{center}

\end{figure}

  The Moon exhibits apsidal precession, which is called Saroscyclus and it  has been known since ancient times. 
  Newton shows in Proposition 43-45 in $Principia$,  that the added force on a single object from a fixed mass
  center which can cause its apsidal precession  must be a central force
  between the planet and a  mass point fixed in space (the Sun). In Proposition 44 he  shows
  that  an inverse-cube force (ICF) might causes the revolving orbits, and  in Proposition 45
  Newton extended his theorem to arbitrary central forces by assuming
  that the particle moved in nearly circular orbit \cite{Chandrasekhar1995}.
  The Moon's  apsidal precession is explained by flattering by the rotating Earth with tide waves, which
  causes an ICF on the Moon. For Newton's  analyse of the Moon's apsidal precession see \cite{Aoki1992}.

  New investigations of isotopes from the Moon reveal that it was created $\approx$ 4.51 billion year ago and 
  $\approx$ 50 to 60 million years after  the emergence of the Earth and our solar system \cite{Barboni2017,Thiemens2019},
  and the Earth contained  the Hadean ocean(s)  with tide waves shortly after the creation of the Moon \cite{Harrison}, 
  so an ICF has
  not affect the overall stability of the Moon's regular orbit. The rotation of the Earth and the Moon's orbit around the Earth results in an ICF which
   has  accelerated the Moon out to its present position with its apsidal precession. The early orbit of the Moon may have had a high eccentricity  \cite{Zuber2006},
    but it is difficult to determine  the evolution of the Moon's orbit  due to the many factors which influence its evolution \cite{Green2017}. One can, however, conclude that the
     presence of an additional force on the Moon due to the tide waves has not affected the overall stability of the Moon's regular orbit.

 The planetary  system and the orbit shown with red in Figure 7 are simulated  with either  ICF or IF included in the attractions.
 The planetary system is affected by including an $\alpha_3 r^{-3}$  ICF, and  the
  systems are destroyed for $\alpha_3 \ge 100$. The ISF planetary system with the planet shown in Figure 7 with red
 contains twenty one planets and only three survived by including $100*r^{-3}$ in the attraction whereas all twenty one planets
 remained in regular orbits for ICF with  $\alpha_3 \le 10*r^{-3}$.
 
 The  orbits in a planetary system with ISF+ICF forces  exhibit the  revolving behaviour predicted
 by Newton: Figure 9 shows the orbit of the planet, also shown in Figure 7, with red without additional attractions, with (green) with  ISF+ICF and with $\alpha_3=-100$,
 and with blue with ISF+IF and with  $\alpha_1=0.01$.  The behaviour of  ISF+ICF and ISF+IF is in agreement with Newton's $Proposition$ $45$.
 Inclusion of IF in the gravitational attractions enhances, however, the revolving behaviour and stabilizes the planetary system, whereas
 inclusion of  the ICF also results in revolving orbits, but it  destabilizes the planetary system.
 The planetary ISF+ICF system is not stable for $\alpha_3 > 100$ and  for pure ICF attractions.

\section{Conclusion}
  The discrete algorithm (Appendix A), derived in \cite{Toxvaerd2022} is used to obtain planetary system with forces other than gravitational forces.
  The main conclusion is, that it is easy to obtain planetary systems with inverse gravitational forces. However, it is not possible to obtain
  planetary systems with inverse cubic gravitational forces, even if one smoothly replaces the inverse square gravitational forces in a stable planetary
  system with inverse cubic forces. A detailed investigation of the planetary system after the  replacement of the forces
  shows, that one can determine a strength
  of the gravitational constant $\beta_3$ for inverse cubic forces for which a planet  either detaches itself
  from the planetary system  for 
  $ \beta_3-\delta$, or are engulfed by the ''Sun" for  $ \beta_3 +\delta$ (Figure 7 and Figure 8).
  So the attractions  in our  universe with inverse square  forces for the gravitational attractions
  between masses and the Coulomb attractions between charges is the limit value
  for regular orbits. A system of objects will,  for inverse attractions with  $\propto r^{-n}$ with  $n \ge 3$, have the well known thermodynamic behaviour
  with  gas-liquid-solid phases, but without regular orbits between units in the system.

  The orbits of the planets in a planetary systems with pure inverse forces have ''revolving orbits". The regular orbits
  deviate, however, significantly  from the   slightly perturbed elliptic orbits in
  an ordinary planetary with additional weak non-gravitational attractions. The principal axis changes with $\approx \pi/3$ at every loops (Figure 1, Figure 2, Figure 6  for the   main part of the regular orbits in a planetary system with inverse forces. But  also changes
   with $\pi/4$ is observed (Figure 3 and Figure 5) together with other smaller, but rather constant changes (Figure 4 and Figure 5). 

 Newton stated in Proposition 43-45 in $Principia$, that the Moons revolving orbits could be explained by an
 additional attraction, $r^{-n}$, to the gravitational attraction with $n \ne 2$. The present simulations of planetary systems with gravitational attractions
 and an additional attractions with either $n=1$ or $n=3$ confirm Newton's Propositions, but whereas attractions with additional inverse attractions
 stabilize the planetary systems, the inclusion of  a weak inverse cubic attractions  also gives '' revolving orbits" (Figure 9), but   it will
 destabilize the planetary
 system by adding sufficient   strong inverse cubic attractions to the inverse square gravitational forces.

\acknowledgments
This work was supported by the VILLUM Foundation’s Matter project, grant No. 16515.
\\
$\textbf{Data Availability Statement}$ Data will be available on request.

\section{Appendix}

The gravitational force, $	\textbf{F}_i(\textbf{r}_i)$, on a planet $i$ at $\textbf{r}_i$ in a planetary systems with $N$ celestial objects  is
\begin{equation}
	\mathbf{F}_i(\textbf{r}_i)= \sum_{j \ne i}^N  \mathbf{F}_{ij}(r_{ij})
\end{equation}	
where the summations over forces $\textbf{F}(r_{ij})$ is given by one of the  Eqn 4-6.

Newton derived the discrete central difference algorithm when he obtained his second law \cite{Toxvaerd2020}.
In Newton's classical discrete dynamics \cite{Newton1687,Toxvaerd2020}  a new  position $\textbf{r}_k(t+\delta t)$ at time $t+\delta t$ of an object
$k$ with the mass $m_k$  is determined by
the force $\textbf{f}_k(t)$ acting on the object   at the discrete positions $\textbf{r}_k(t)$  at time $t$, and  
 the position $\textbf{r}_k(t-\delta t)$ at $t - \delta t$  as
 \begin{equation}
	 	 m_k\frac{\textbf{r}_k(t+\delta t)-\textbf{r}_k(t)}{\delta t}
		 			=m_k\frac{\textbf{r}_k(t)-\textbf{r}_k(t-\delta t)}{\delta t} +\delta t \textbf{f}_k(t),	
					 \end{equation}
					 where the momenta $ \textbf{p}_k(t+\delta t/2) =  m_k (\textbf{r}_k(t+\delta t)-\textbf{r}_k(t))/\delta t$ and
					  $  \textbf{p}_k(t-\delta t/2)=  m_k(\textbf{r}_k(t)-\textbf{r}_k(t-\delta t))/\delta t$ are constant in
					  the time intervals in between the discrete positions.
					  Newton postulated Eq. (A2) and obtained his  second law, and the analytic dynamics  in the  limit $ lim_{\delta t \rightarrow 0}$.

					   The algorithm, Eq. (A2), is usual presented  as the ''Leap frog" algorithm for the velocities
					   \begin{equation}
						   \textbf{v}_k(t+\delta t/2)=  \textbf{v}_k(t-\delta t/2)+ \delta t/m_k  \textbf{f}_k(t).
					   \end{equation}
					   The positions
					   are determined from the discrete values of the momenta/velocities as
					   \begin{equation}
						   \textbf{r}_k(t+\delta t)= \textbf{r}_k(t)+ \delta t \textbf{v}_k(t+\delta t/2).	  
					   \end{equation}	  
Let all the spherically symmetrical objects
 have the same (reduced)  number density $\rho= (\pi/6)^{-1} $ by which
  the diameter $\sigma_i$ of the spherical object $i$ is 

   \begin{equation}
	   	 		 \sigma_i= m_i^{1/3}
				  \end{equation}
				    and   the collision diameter 
				    \begin{equation}
					    	\sigma_{ij}=	\frac{\sigma_{i}+\sigma_{j}}{2}.
				    \end{equation}	
				     If  the distance $r_{ij}(t)$ at time $t$ between two objects is less than $\sigma_{ij}$ 
				     the two objects merge to one spherical symmetrical object with mass

				     \begin{equation}
					     m_{\alpha}= m_i + m_j,
				     \end{equation}	 
				     and diameter
				     \begin{equation}
					      \sigma_{\alpha}= (m_{\alpha})^{1/3},
				     \end{equation}
				     and with the new object $\alpha$  at the position
				     \begin{equation}
					     \textbf{r}_{\alpha}(t)= \frac{m_i}{m_{\alpha}}\textbf{r}_i(t)+\frac{m_j}{m_{\alpha}}\textbf{r}_j(t),
				     \end{equation}	
				     at the center of mass of the the two objects before the fusion.
				     (The   object $\alpha$ at the center of mass of the two merged objects $i$ and $j$ might occasionally be near another object $k$
				     by which more objects merge, but after the same laws.)

The momenta  of the objects in the discrete dynamics just before the fusion are $\textbf{p}^N(t-\delta t/2)$ and the
total momentum of the system is conserved  at the fusion if
\begin{equation}
	\textbf{v}_{\alpha}(t-\delta t/2)= \frac{m_i}{m_{\alpha}}\textbf{v}_i(t-\delta t/2)+ \frac{m_j}{m_{\alpha}}\textbf{v}_j(t-\delta t/2),
\end{equation}
which determines the  velocity $\textbf{v}_{\alpha}(t-\delta t/2)$ of the merged object.

The algorithm for planetary system consists of the equations (A3)+(A4) for time steps without merging of objects, and the fusion of objects
is given by the equations (A6),(A7), (A8), (A9) and (A10).

Newtons discrete algorithm (A3), which is used in almost all
MD simulations, is usually called the Verlet- or Leap-frog algorithm and it  has the same invariances as his exact analytic dynamics \cite{Toxvaerd2022,Toxvaerd1994,Toxvaerd2012}.
The invariances are maintained by the extension to planetary systems (A6),(A7), (A8), (A9) and (A10) \cite{Toxvaerd2022} .

 The gravitational strengths in the article  are in units of $\beta_i^*=G=1$ and the mass $m_i(0)=1$ and diameters
 of the planets $\sigma_i(0)=1$
at the start time $t=0$. For units and set-up of the systems see also \cite{Toxvaerd2022}. The planetary systems in the articles are obtained for thousand objects,
which at  $t=0$ are separated with a mean distance $<r_{ij}> \approx 1000$ and with a Maxwell-Boltzmann distributed velocities with  mean velocity $ <v_i> \approx 1$, for the  set-up of the systems see also \cite{Toxvaerd2022}. 
The systems are followed at least $10^9$ MD time steps, i.e. $t=2.5\times 10^6$ time-units, which
 corresponds to $\approx 10^3$ to $10^4$ orbits for a planet.

\end{document}